\documentclass[fleqn,usenatbib]{mnras}

\usepackage[T1]{fontenc}

\DeclareRobustCommand{\VAN}[3]{#2}
\let\VANthebibliography\thebibliography
\def\thebibliography{\DeclareRobustCommand{\VAN}[3]{##3}\VANthebibliography}

\usepackage{subcaption}
\usepackage{graphicx}	
\usepackage{amssymb}
\usepackage{amsmath}	
\usepackage{enumitem}
\usepackage{xcolor}
\usepackage{xspace}

\usepackage{adjustbox}

\title[Low-frequency wideband timing of InPTA pulsars observed with the uGMRT]{Low-frequency wideband timing of InPTA pulsars observed with the uGMRT}

\author[Nobleson et al.]{K Nobleson$^1$,\thanks{E-mail:nobleson.phy@gmail.com}
Nikita Agarwal$^{2}$,
Raghav Girgaonkar$^3$,
Arul Pandian$^{4}$,
\newauthor
Bhal Chandra Joshi$^5$,
M A Krishnakumar$^{6}$,
Abhimanyu Susobhanan$^{5,7}$,
Shantanu Desai$^{3}$,
\newauthor
T Prabu$^{4}$,
Adarsh Bathula$^{8}$,
Timothy T Pennucci$^9$,
Sarmistha Banik$^{1}$,
Manjari Bagchi$^{10,11}$,
\newauthor
Neelam Dhanda Batra$^{12}$,
Arpita Choudhary$^{10}$,
Subhajit Dandapat$^{7}$,
Lankeswar Dey$^{7}$,
\newauthor
Yashwant Gupta$^5$,
Shinnosuke Hisano$^{14}$,
Ryo Kato$^{15}$,
Divyansh Kharbanda$^{3}$,
\newauthor
Tomonosuke Kikunaga$^{14}$,
Neel Kolhe$^{16}$
Yogesh Maan$^5$,
Piyush Marmat$^{17}$,
P Arumugam$^{17}$,
\newauthor
P K Manoharan$^{18}$,
Dhruv Pathak$^{10,11}$,
Jaikhomba Singha$^{17}$,
Mayuresh P Surnis$^{19}$,
\newauthor
Sai Chaitanya Susarla$^{20}$,
Keitaro Takahashi$^{21,22}$.\\
$^{1}$Department of Physics, BITS Pilani, Hyderabad Campus, Hyderabad 500078, Telangana, India\\
$^{2}$Department of Electronics and Communication Engineering, Manipal Institute of Technology, Manipal Academy of Higher Education, \\ Manipal 576104, Karnataka, India\\
$^{3}$Department of Physics, Indian Institute of Technology Hyderabad, Kandi, Telangana 502285, India\\
$^{4}$Raman Research Institute, Bengaluru 560080, Karnataka, India \\
$^{5}$National Centre for Radio Astrophysics, Tata Institute of Fundamental Research, Pune 411007, Maharashtra, India\\
$^{6}$Fakult{\"a}t f{\"u}r Physik, Universit{\"a}t Bielefeld, Postfach 100131, 33501 Bielefeld, Germany\\
$^{7}$Department of Astronomy and Astrophysics, Tata Institute of Fundamental Research, Mumbai 400005, Maharashtra, India\\
$^{8}$Department of Physics, Indian Institute of Science Education and Research Mohali, Sector 81, Punjab 140306, India\\
$^{9}$Institute of Physics, Eötvös Loránd University, Pázmány P.s. 1/A, 1117 Budapest, Hungary\\
$^{10}$The Institute of Mathematical Sciences, 4th Cross Road, CIT Campus, Taramani, Chennai 600 113, India \\
$^{11}$Homi Bhabha National Institute, Training School Complex, Anushakti Nagar, Mumbai 400094, Maharashtra, India\\
$^{12}$Department of Physics, Indian Institute of Technology Delhi, New Delhi-110016, India\\
$^{14}$Kumamoto University, Graduate School of Science and Technology, Kumamoto, 860-8555, Japan\\
$^{15}$Osaka City University Advanced Mathematical Institute, 3-3-138, Sugimoto, Sumiyoshi-ku, Osaka, 558-8585, Japan\\
$^{16}$Department of Physics, St. Xavier's College (Autonomous), Mumbai 400001, Maharashtra, India\\
$^{17}$Department of Physics, Indian Institute of Technology Roorkee, Roorkee 247667, Uttarakhand, India\\
$^{18}$Arecibo Observatory, University of Central Florida, Arecibo, PR 00612, USA\\
$^{19}$Jodrell Bank Centre for Astrophysics, Department of Physics and Astronomy, The University of Manchester, Manchester, M13 9PL, UK\\
$^{20}$Center for Astronomy, Department of Physics, National University of Ireland Galway, Galway H91 TK33, Ireland\\
$^{21}$International Research Organization for Advanced Science and Technology, Kumamoto University, 2-39-1 Kurokami, \\ Kumamoto 860-8555, Japan\\
$^{22}$Department of Physics, Kumamoto University, 2-39-1, Kurokami, Kumamoto 860-8555, Japan\\
}

\date{Accepted XXX. Received YYY; in original form ZZZ}

\begin{document}
\label{firstpage}
\pagerange{\pageref{firstpage}--\pageref{lastpage}}
\maketitle


\begin{abstract}
High-precision measurements of the pulsar dispersion measure (DM) are possible using telescopes with low-frequency wideband receivers. We present an initial study of the application of the wideband timing technique, which can simultaneously measure the pulsar times of arrival (ToAs) and DMs, for a set of five pulsars observed with the upgraded Giant Metrewave Radio Telescope (uGMRT) as part of the Indian Pulsar Timing Array (InPTA) campaign. We have used the observations with the 300 -- 500~MHz band of the uGMRT for this purpose.
We obtain high precision in DM measurements with precisions of the order $10^{-6}$ cm$^{-3}$~pc. The ToAs obtained have sub-$\mu$s precision and the root-mean-square of the post-fit ToA residuals are in the sub-$\mu$s range. We find that the uncertainties in the DMs and ToAs obtained with this wideband technique, applied to low-frequency data, are consistent with the results obtained with traditional pulsar timing techniques and comparable to high-frequency results from other PTAs. This work opens up an interesting possibility of using low-frequency wideband observations for precision pulsar timing and gravitational wave detection with similar precision as  high-frequency observations used conventionally. 

\end{abstract}

\begin{keywords}
stars: pulsars: InPTA: gravitational waves 
\end{keywords}



\section{Introduction}  \label{sec:intro}
Pulsars are rotating neutron stars, which emit pulsed radiation, observed mainly in radio wavelengths. This radiation traverses the interstellar medium (ISM), where it gets dispersed due to the presence of free electrons, causing a delay in the times of arrival (ToAs) of pulses as a function of frequency \citep{2012hpa..book.....L}. This dispersion delay is characterized by the dispersion measure (DM), which is proportional to the cumulative column density of free electrons in the interstellar medium. One can get precise measurements of the DM by measuring the pulse ToAs simultaneously at different frequencies.

The increasing  availability of wideband receivers and backends presents new opportunities for high precision timing with wideband timing (WT) techniques \citep[][] {2014ApJ...790...93P, 2014MNRAS_443_3752L}. Such high precision measurements are very important for pulsar timing arrays \citep[PTAs:][]{1990ApJ...361..300F}, such as the Parkes Pulsar Timing Array \citep[PPTA:][]{hobbs2013,krh+2020}, the European Pulsar Timing Array \citep[EPTA:][]{kc2013,dcl+2016}, the North American Nanohertz Observatory for Gravitational Waves \citep[NANOGrav:][]{McLaughlin2013,Alam21}, the Indian Pulsar Timing Array \citep[InPTA:][]{jab+2018} and the International Pulsar Timing Array (IPTA) consortium, which combines the data and resources from various PTA \citep{pdd+19} experiments in order to search for nanohertz gravitational waves (GWs). Such a technique not only  provides high precision ToAs, but also yields simultaneous estimates of the DM variations for the millisecond pulsars (MSPs) being timed. The wideband technique has been applied to 
datasets such as the NANOGrav 12.5-year dataset \citep[][]{Alam21}. 
In this work, we describe the application of this technique to low frequency (below 400 MHz) observations for the first time, which is complementary to the recent application in the  400-800 MHz frequency range using data from CHIME and GBT-L~\citep{Fonseca21}.

Pulsars are bright at frequencies below 1~GHz. The higher signal to noise ratio (S/N) of MSPs at these frequencies could potentially yield higher precision ToAs. However, the electron distribution in the ISM has a dominant effect on the pulse shape and arrival time of the pulsed signal from these stars at such low frequencies \citep{Cordes_2016}. Scattering due to the ISM not only broadens the pulsed signal, but also delays the pulse by a factor roughly proportionate to its width, leading to inaccurate timing measurements  \citep{Levin_2016}. Additionally, the time variability of the DM due to the dynamic nature of the ISM introduces a correlated noise \citep[][]{2017MNRAS.464.2075S} in the GW analysis. This ISM noise is slowly varying and is covariant with the signature of the stochastic gravitational wave background (SGWB), formed by an incoherent superposition of GWs coming from an ensemble of supermassive black hole binaries \citep{romano2017detection}. As the magnitude of this noise is much larger at low frequencies, PTAs have conventionally used higher frequency observations for the search of nanohertz GWs, despite a higher S/N at low radio frequencies. 

As the ISM noise affects higher frequency observations as well, PTA experiments typically correct high frequency ToAs using DM estimates obtained from quasi-simultaneous narrow band observations at two or three widely separated observing frequencies \citep{You.2007,2013MNRAS.429.2161K, 2021ApJS..252....4A}. The alignment of the fiducial point of the pulse at different observing frequencies is critical in such measurements. This can introduce a systematic bias in the  measured DMs as well as in other pulsar timing parameters~\citep{lkd+17}. Furthermore, extreme scattering events \citep{10.1093/mnras/stx3101}, DM events \citep{Lam_2018}, and profile changes \citep{ssj+21,2021ATel14642....1X, 2021ATel14652....1M} have been reported in some of the pulsars, such as PSR J1713+0747, which complicate the GW analysis of the PTA data.

Alternatively, wideband receivers have been employed between 700 $-$ 4000~MHz by PPTA \citep{Johnston:2021jgc} for higher precision DM measurements. Application of wideband techniques to such data provide a robust way to correct the profile evolution with frequency as well as ISM noise, including the corruption of data by abrupt ISM events. However, the dispersive delay due to the ISM varies as $\nu^{-2}$, whereas the pulse scatter broadening evolves as $\nu^{-4.4}$, if a Kolmogorov turbulence is posited for the ISM \citep{doi:10.1146/annurev.aa.15.090177.002403, cordes1986refractive}. While this strong frequency dependence is challenging for low frequency PTA observations, application of the wideband techniques to observations between 300 $-$ 800 MHz can, in principle, better account for these effects and can provide very precise ToAs. Thus, the application of this technique to frequencies between 300 to 800 MHz promise to make low radio frequency PTA observations as useful as high frequency observations.

InPTA uses wideband coherently dedispersed observations with the 300 $-$ 500 MHz band of the upgraded GMRT \citep[uGMRT:][] {gak+17}. In this paper, we provide a proof of principle application of wideband timing technique {\tt PulsePortraiture} \footnote{https://github.com/pennucci/PulsePortraiture} \citep{2016ascl.soft06013P} to such observations for five pulsars. This complements the DM measurements by both the methods reported in ~\citet{kmj21} computed using the {\tt DMcalc} package. This paper is structured as follows. InPTA observations used in this paper are briefly described in Section \ref{sec:obs} and a description of analysis of these data using {\tt PulsePortraiture}, in Section \ref{sec:WBanaly}. The results obtained using this method for five pulsars are presented in Section \ref{sec:result}. The paper concludes with a discussion in Section~\ref{sec:conc}.

\section{Observations and data reduction} \label{sec:obs}

The InPTA collaboration has been monitoring five pulsars (PSRs J1643$-$1224, J1713+0747, J1909$-$3744, J1939+2134, and J2145$-$0750) since 2018 with a cadence of around 15 days, using the uGMRT. These observations were carried out by dividing uGMRT's 30 antennas into two sub-arrays. These pulsars are observed in Band 3 (300-500 MHz) and Band 5 (1060-1460 MHz)  simultaneously using separate sub-arrays. The nearest 10 central square antennas were included in the Band 3 sub-array, where the data  were coherently dedispersed in real-time \citep[]{de2016real} and recorded using the GMRT Wideband Backend \citep[GWB:][]{Reddy+2017} with a 200 MHz band-pass. The number of sub-bands for the recording vary between 64 to 1024 with the sampling time used ranging from 5 to 40 $\mu$s, respectively. The observation time per pulsar is typically about 55 minutes. 

The recorded data were reduced offline using an automated data reduction pipeline {\tt pinta} \citep{pinta}, developed for the InPTA data. Using the known pulsar ephemeris \citep[IPTA DR2:][]{pdd+19}, this pipeline partially folds the data from all the sub-bands  into sub-integrations of 10-second duration to archive files in the {\tt TIMER} format \citep[][]{van_straten_bailes_2011}. Radio-frequency interference mitigation was performed by {\tt RFIClean} \citep{Maan+2020}. Before our analysis, all the  reduced data were further collapsed in time to a single integration, with 64 sub-bands.  In this work, we report the results for the aforementioned five pulsars 
observed in Cycle 39 of uGMRT covering the period between November 2020 to April 2021. Since our goal is to demonstrate the application of WT at low frequencies, only Band 3 data were used for the work presented here. The selection of these pulsars was made based on the different pulse morphology and a range of observed scatter-broadening in the pulse profiles. PSR J1909--3744 shows systematic changes in DM but also has the best achievable timing solution. PSRs J1643--1224 \citep{10.1093/mnras/stx580} and J1939+2134 \citep{Ramachandran_2006} show large systematics due to a  variation in the pulse scatter broadening. PSRs J1713+0747 \citep{Dolch_2014} and J2145--0750 \citep{2004A&A...426..631L} are bright pulsars in our sample, which also show scatter broadening, profile evolution, and scintillation in our frequency range apart from epoch to epoch DM variations. Amongst these pulsars, J1909--3744 and J2145--0750 are the best timed pulsars. This diversity of frequency dependent effects in this sample of pulsars is useful to evaluate the efficacy  and the systematic errors in the wideband technique. However, in future, we shall apply this technique to all the pulsars observed by InPTA, using the  full observational data.

\section{Wideband analysis} \label{sec:WBanaly}
We now provide some technical details of the methodology used in the {\tt PulsePortraiture} package, which is the workhorse behind the WT analysis used in this work. More details can be found in ~\citet{2014ApJ...790...93P,Pennucci19}. The wideband data processing can be divided into three stages using the following modules: {\tt ppalign}, {\tt ppspline}, and {\tt pptoas}. We now describe each of these modules below.

\begin{itemize}
    \item {\tt ppalign}:
The first step in {\tt PulsePortraiture} involves creating  a two-dimensional template, containing the  pulse amplitude as a function of the frequency and phase. This step is done in the {\tt ppalign} module.
Each phase-frequency sub-integration in the data set is called a data ``portrait''. 
The starting point in constructing this portrait is nearly the same as in the traditional analysis, which consists of an iterative procedure of co-adding all the significant total intensity profiles  at all frequencies in a given band~\citep{Demorest2007}. The only difference is that instead of aligning each data portrait compared to a constant profile portrait using only a phase shift, each profile is rotated by a factor proportional to the inverse-square of its frequency. By doing so, we can minimize the dispersive delays caused by DM changes, which could smear the average portrait. This iterative process is carried out multiple times to create a final average portrait. 
Regardless of averaging, the choice of alignment will be covariant with the absolute DM. In our analysis, we used a single epoch data to construct the portrait since the signal to noise ratio was reasonably high and to ensure, when comparing with {\tt DMcalc}, that the fiducial DM in both the methods are same, minimizing the offset between the DM estimations in the two methods.  
\\
\item {\tt ppspline}: 
This module does the principal component analysis (PCA) decomposition of the average portrait, followed by reconstruction of the template profile based on the significant eigenvectors.
We apply PCA to the  average portrait profiles, whose  dimensions are  $n_{chan} \times n_{bin}$, where $n_{chan}$ are the total number of frequency bins and $n_{bin}$ are the total number of phase bins, which encompass the observed bandwidth and pulsar profile, respectively. Unlike conventional PCA, we do not select the optimum basis vectors (here referred to as eigenprofiles) based on the largest eigenvalues, in order to avoid getting contaminated by radiometer noise. Instead, we choose the top 10 eigenprofiles ranked according to their S/N. In order to determine the S/N, the mean profile as well as the eigenprofiles were first smoothened using Stationary Wavelet Transform based denoising, and the S/N was then calculated using the definition in \citet{Arzu15}.

The mean-subtracted profiles are projected onto each of the eigenprofiles to obtain a set of coordinate coefficients. A low-degree spline function is fitted to these coefficients, which is parameterized by frequency and encompasses the evolution of the pulse profile shape.  By linearly combining the eigenprofiles $e_i$ using the spline coefficients $B_i$ and adding it to the mean profile $\tilde{p}$, a template profile $T(\nu)$ at any frequency $\nu$ can be created as follows:

\begin{equation}
T(\nu) = \sum_{i=1}^{n_{eig}} B_i(\nu) \hat{e_i} + \tilde{p}
\end{equation}
\\
\item {\tt pptoa}:
In this step, the DMs and ToAs are calculated.  The ToA and DM pair from each observation are obtained by minimizing the $\chi^2$ value as follows~\citep{2014ApJ...790...93P},
\begin{equation}
\label{eqn:wb_toa}
\chi^2 = \sum_{n,k} \frac{|d_{nk} - a_n t_{nk}e^{-2\pi i k \phi_n}|^2}{\sigma_n^2}.
\end{equation}
Equation~\ref{eqn:wb_toa} takes the same form as the conventional ToA likelihood used in the Fourier-domain phase-gradient shift algorithm \citep{Taylor92}, except here there is an additional index $n$, which labels the frequency channel with center frequency $\nu_n$.  $k$ is the index corresponding to the Fourier frequency, which is conjugate to the rotational phase or time.  The other terms in the equation are as follows: $d_{nk}$ is the discrete Fourier transform of the data profiles, $\sigma_n^2$ are their corresponding Fourier domain noise levels, $t_{nk}$ is the discrete Fourier transform of the template profiles, $a_n$ are the scaling amplitudes for each template profile, and $\phi_n$ are the phase shifts applied to each template profile.
The two fitted parameters of interest, $\phi_0$ (which corresponds to the ToA) and the DM, arise because the phase offsets $\phi_n$ for each profile are constrained to follow the cold-plasma dispersion law:
\begin{equation}
\phi_n(\nu_n) = \phi_0 + \frac{K \times DM}{P_s} (\nu_n^{-2}-\nu_{\phi_0}^{-2}),
\end{equation}
where $P_s$ is the instantaneous spin period of the pulsar, $K$ is the dispersion constant ($4.148808 \times$ $10^3$ MHz$^2$ cm$^3$ pc$^{-1}$s).
The ToAs and DMs are simultaneously fit for in such a way that there is zero covariance between them. The ToAs and DMs are then written to a file.

\end{itemize}

\begin{figure*}
\includegraphics[scale =0.4]{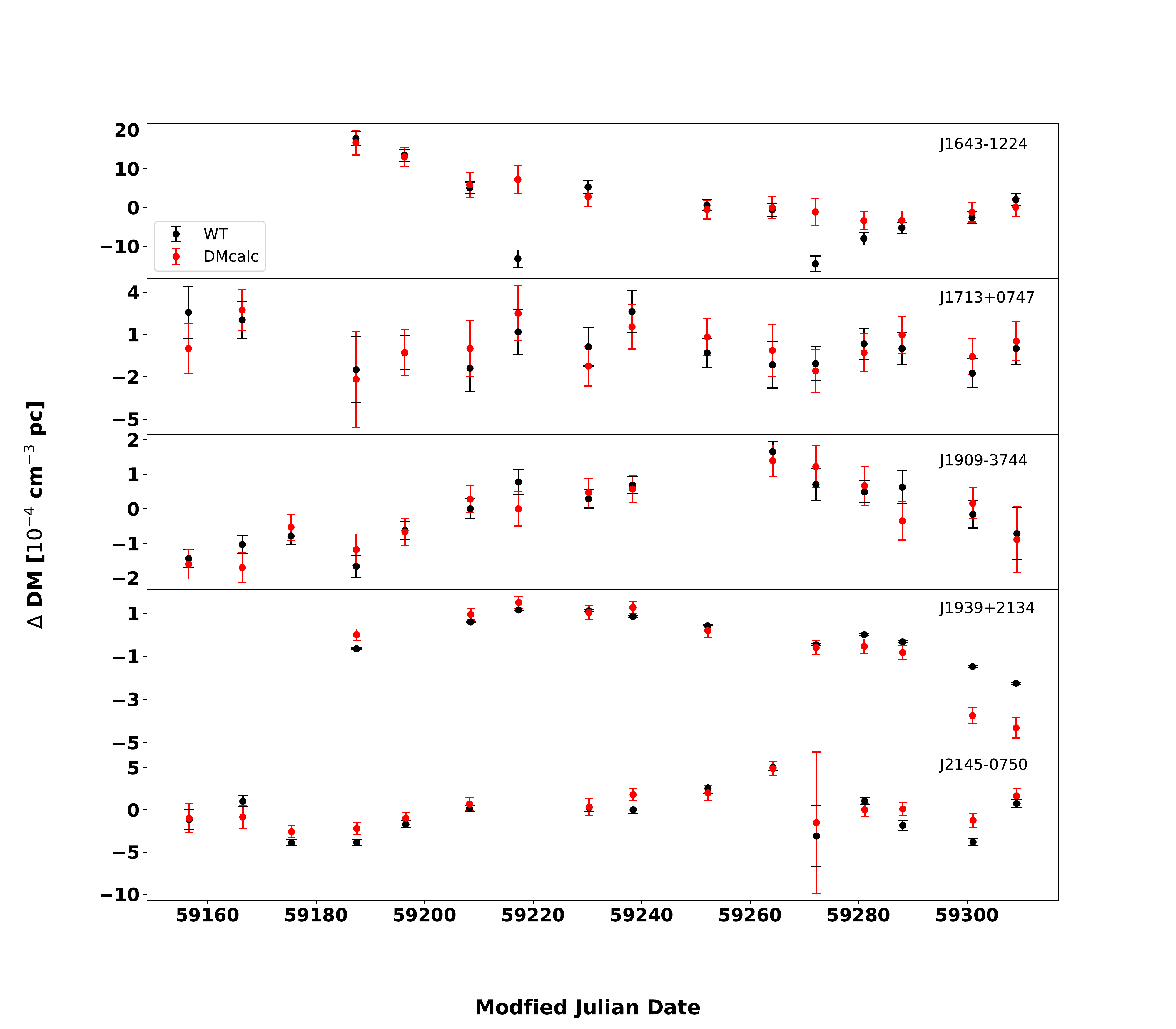}
\caption{The median-subtracted DM variations are plotted for the pulsars in our sample observed at uGMRT (400 MHz with 200 MHz bandwidth) from November 2020 to April 2021. The black points correspond to WT and the red points correspond to {\tt DMcalc}. The median DM uncertainties range from 3$\times 10^{-6}$ to 1$\times 10^{-4}$~cm$^{-3}$~pc. The median DM values and corresponding uncertainties for each pulsar are listed in the Table \ref{tab:obs}.} 
\label{fig:DM}
\end{figure*}

\begin{figure*}
\includegraphics[scale =0.4]{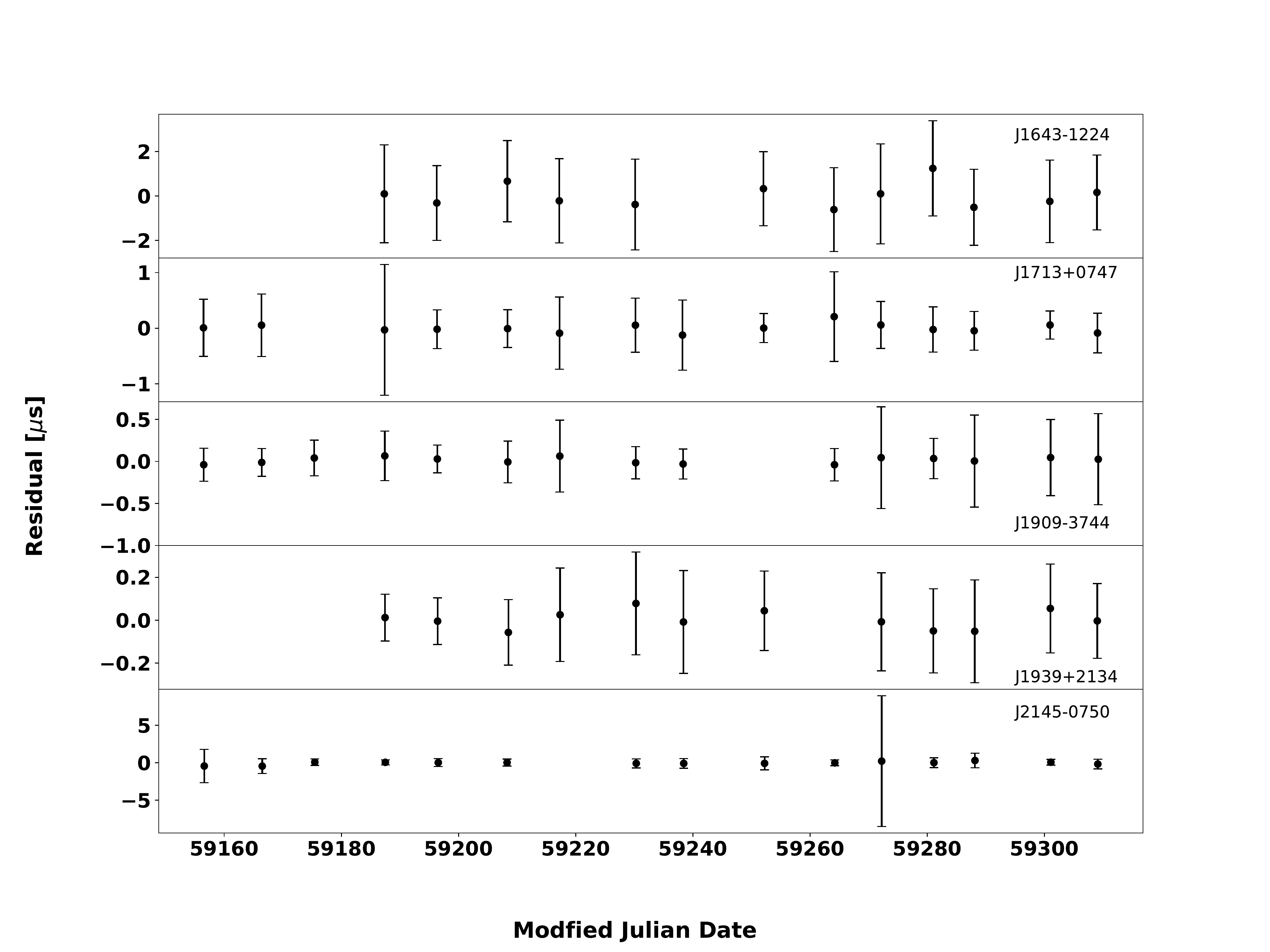}
\caption{The timing residuals are plotted for the pulsars in our sample observed at uGMRT (400 MHz with 200 MHz bandwidth) from November 2020 to April 2021. The first three points in PSRS J1643$-$1224 and J1939$+$2134 are missing because of non-detection of pulsars in these epochs. The median ToA uncertainties range from 0.20 to 1.87 $\mu$s. The ToA uncertainty and postfit rms for each pulsar are listed in the Table \ref{tab:obs}.}
\label{fig:res}
\end{figure*}

A likelihood that is implemented in the pulsar timing software package {\tt TEMPO} \citep{2015ascl.soft09002N} effectively uses the wideband DM measurements from the ToAs as priors on the DM model parameters. Details of this procedure can be found in Appendix B of \citet{Alam21}. Therefore, the wideband analysis of InPTA data simultaneously yields a DM timeseries as well as the residuals with respect to the fitted ephemeris. These are shown in Figures \ref{fig:DM} and \ref{fig:res}. The spin parameters, $F_0$ and $F_1$, and the orbital Keplerian parameters, namely the orbital period (P$_b$), the projected pulsar semi-major axis ($a_p \sin i$), longitude and the epoch of periastron passage ($\omega$, $T_0$) are fitted. As we do not expect to improve the positional parameters of the pulsar over the six-month time-span of our data, these are not fitted. We also did not fit the solar wind model. The TDB \citep{1988A&A...194..304G} coordinates with DE436 \citep{folkner2016jpl} ephemeris are used throughout the analysis. The parameter files used in this analysis are taken from the NANOGrav 12.5-year data release \citep[NG:][]{Alam21}. The results of these analyses are presented in the next section.

\begin{figure*}
\includegraphics[scale =0.6]{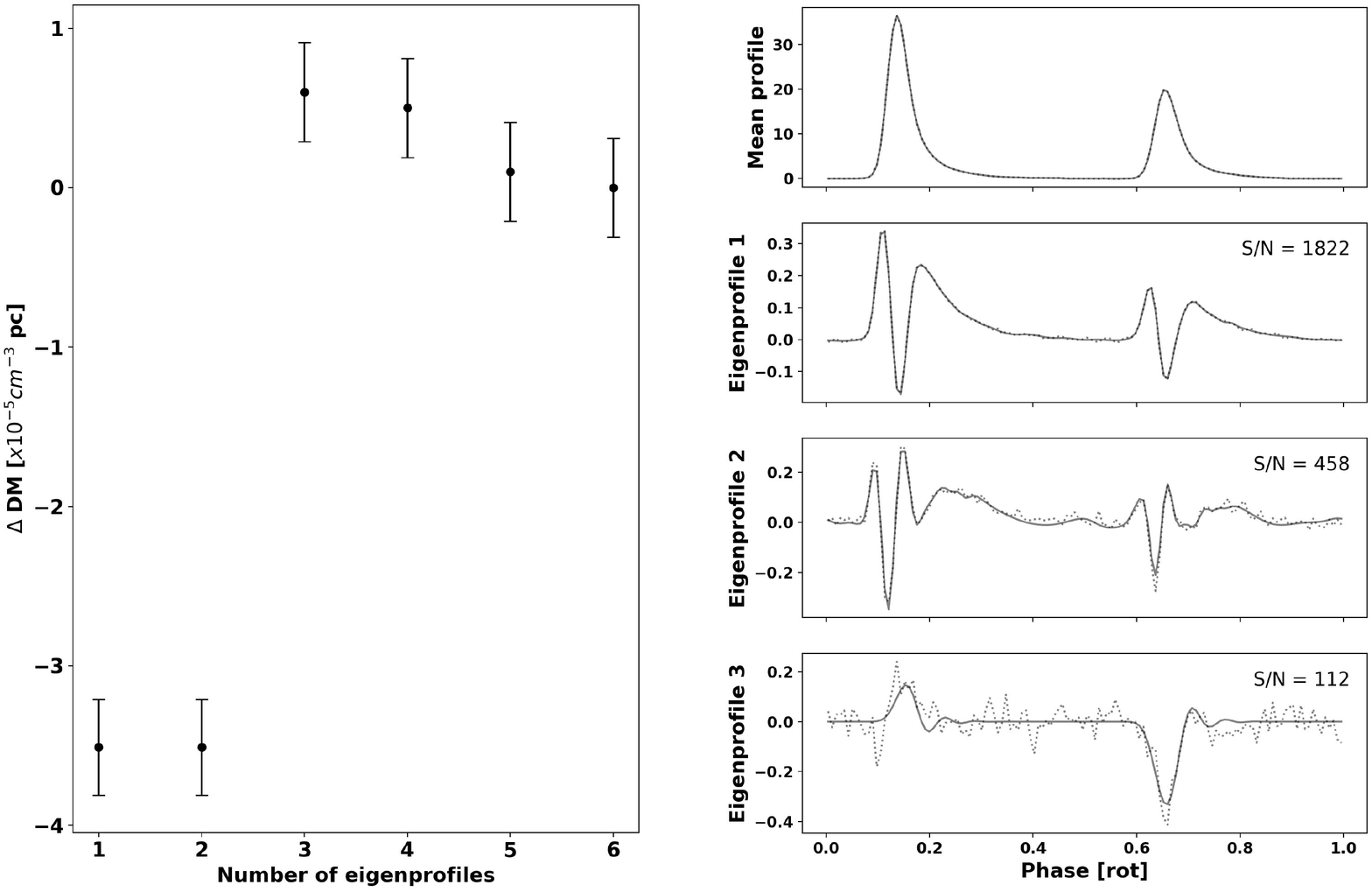}
\caption{The plot on the left depicts how the DM estimation varies with the number of eigenprofiles chosen for J1939$+$2134. On the Y-axis, difference in DM (each result is subtracted from the DM estimate derived using six eigenprofiles) is plotted. A portrait is created using the data from the template epoch. {\tt ppspline} is used to process the portrait by varying the number of eigenprofiles. The DM estimate and the DM uncertainty for each variation is then obtained after processing with {\tt pptoas}. With three or more eigenprofiles, the DM estimates tend to converge. The profile evolution was modelled using three eigenprofiles, the others are presented for illustration purposes. The error bars correspond to the uncertainty in the DM measurement. The mean profile and the eigenprofiles corresponding to the PCA decomposition of PSR J1939$+$2134 data are presented for visual reference on the right panel. The top three highest S/N profiles are shown here. The median DM is 71.017317. The grey points represent the data's computed values, while the dark lines are the smoothed curves that comprise part of the model.}
\label{fig:pca}
\end{figure*}

\begin{figure*}
\includegraphics[scale =0.35]{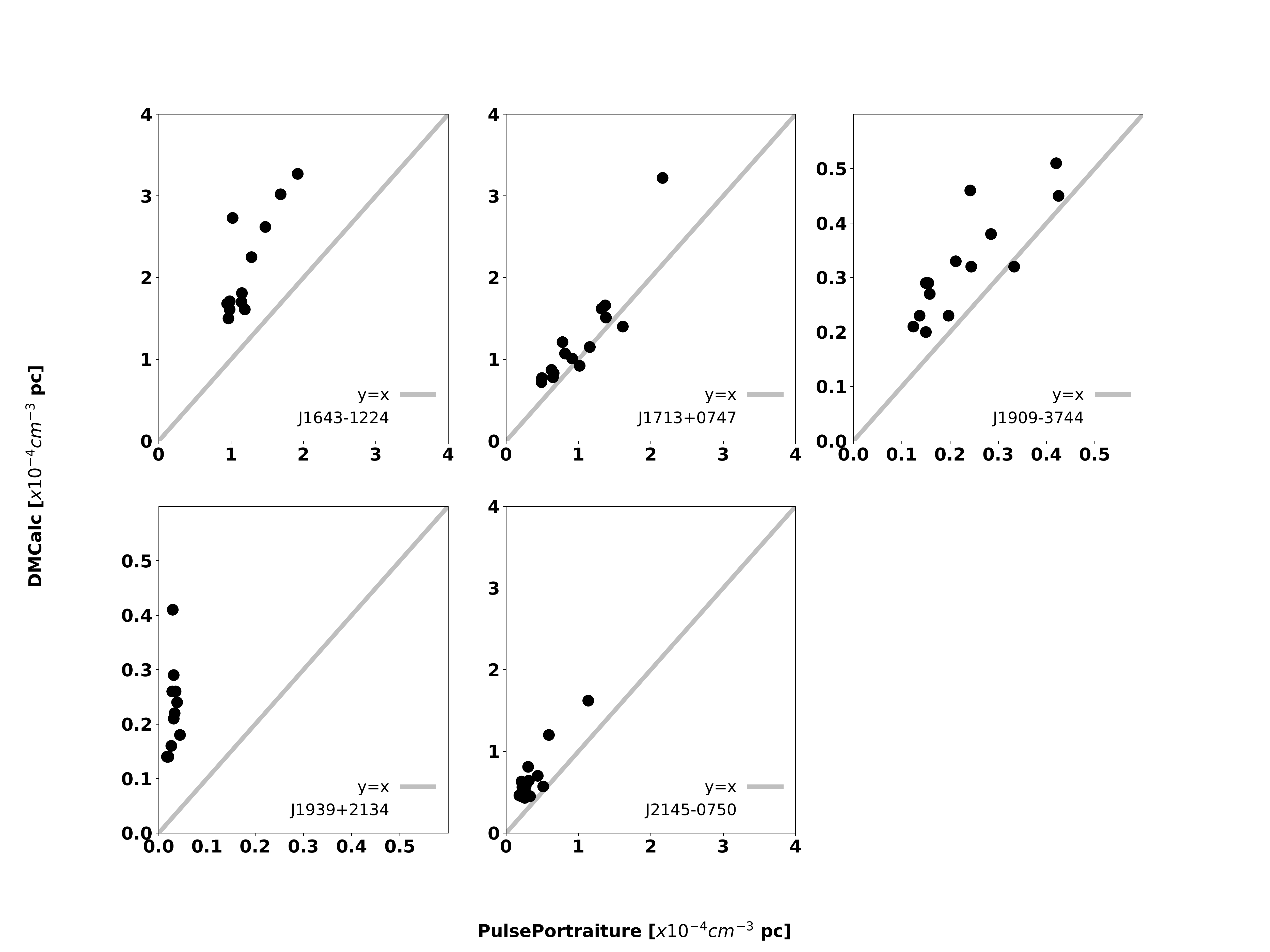}
\caption{This plot shows the comparison of the DM uncertainties obtained with {\tt DMcalc} and {\tt PulsePortraiture} for the pulsars in our study. The uncertainty from both the methods are consistent except for J1939$+$2134 and J1643--1224. The discrepancy could be due to the fact that we have not taken scattering effects into account and these pulsars show significant variation in pulse broadening at low frequencies.}
\label{fig:uncert}
\end{figure*}

\section{Analysis and Results} \label{sec:result}
The main goal of this work is to measure the wideband ToAs and DMs for the low-frequency pulsar data in order to study its suitability for PTA experiments. We applied this technique to a sample of five pulsars, which were  observed by InPTA, and obtained DM estimates as well as  timing residuals.  

In a nutshell, Figure~\ref{fig:DM} displays the DM estimations from WT and a comparison to DM values obtained using  {\tt DMcalc} {\citep[][]{kmj21}}. The WT DM estimates obtained by measuring each WT ToA are plotted here. Figure~\ref{fig:res} demonstrates the WT ToAs with a precision less than 1.9 ${\mu}$s and rms post-fit residuals of 0.5 ${\mu}$s or better. The eigenprofiles corresponding to the PCA decomposition of the PSR J1939$+$2134 data are shown in Figure~\ref{fig:pca}. The DM uncertainties obtained using WT and {\tt DMcalc} methods are compared in Figure~\ref{fig:uncert}. Table~\ref{tab:obs} summarises the results of WT method, whereas Table~\ref{tab:par} compares the WT and narrow-band timing (NT) solutions.

The DM estimations were done by independent WT analyses with varying the number of eigenprofiles, the number of bins, as well as the epoch data used to make the average portrait, to understand their effect on the results.
We draw the following conclusions based on these tests:
\begin{enumerate}
\item  In Figure~\ref{fig:pca} (left panel), we show how the DM estimates correlate with the number of eigenprofiles for PSR J1939$+$2134. With one and two eigenprofiles, the DM value is underestimated. The first two eigenprofiles span only about {94\%} of the profile evolution. The first, second, and third eigenprofiles together span about {99\%} of the profile evolution. When the number of eigenprofiles is three or above, the DM values are consistent within the error bars. On the right panel, the mean profile and eigenprofiles corresponding to the PCA decomposition of PSR J1939$+$2134 data are plotted for visual reference. The grey points are the computed values from the data and the dark lines are the smoothed curves that comprise the model. Based on similar analysis, the optimum number of eigenprofiles, required for accurate DM estimations for the five pulsars in our sample, are as follows: For PSRs J1643$-$1224, J1713$+$0747, and J1939$+$2134 the DM estimates with $n_{eig}$ = 3 or more are consistent with each other. Therefore, DM estimation can be done with a minimum of three eigenprofiles. For J1909$-$3744 and J2145$-$0750, the DM estimates with $n_{eig}$ = 2 or more are consistent within the error bars. Therefore, for these pulsars a minimum of $2$ eigenprofiles are required.    
\item  The DM uncertainties are smaller with larger number of bins, compared to those with smaller number of bins. This is consistent across all the five pulsars in our sample.  
\item The DM estimates obtained using the different averaged portraits have an offset among them. However, the median subtracted DMs are consistent with each other within the error bars. This is consistent across all the five pulsars in our sample.
\end{enumerate}

For our WT analysis, we reduced the data to 64 channels, and one sub-integration for all the pulsars in our sample. PSRs J1643--1224, J1713$+$0747, and J1909--3744, data have 256 bins; J1939+2134 data have 128 bins; and J2145--0750 data have 1024 bins. The WT results for all the pulsars are summarized in Table~\ref{tab:obs}.


Conventional NT ToAs and DM estimation were performed with 4 subbands for J1643--1224, and 16 subbands for J1713$+$0747, J1909--3744, J1939$+$2134 and J2145--0750.
Frequency-resolved\footnote{A single pulsar has multiple smoothed templates spanning the bandpass to generate ToAs at respective frequencies.} templates were created for each pulsar using a wavelet smoothing algorithm \citep{Demorest2013}, implemented as the \texttt{psrsmooth} command in \texttt{PSRCHIVE} \citep{hotan_van_straten_manchester_2004}, on the same epoch data with the same number of bins as used for the wideband templates.
These templates were aligned using the same fiducial DMs as the ones used to align the wideband templates. 
The ToAs were computed from the frequency-resolved profiles using the Fourier Phase Gradient algorithm \citep{Taylor1992} available in the \texttt{pat} command of \texttt{PSRCHIVE}.
The resulting ToAs were then fitted for the DM, spin-down parameters F0 and F1, and the binary parameters PB, A1 and T0/TASC (where applicable) using \texttt{TEMPO2}.
In addition, the epoch-by-epoch DM variations were modeled by fitting for the `DMX' parameters in the pulsar ephemeris.
DMX is a piecewise-constant representation of the DM variability that is included in the timing model. 
A separate DM is estimated for each DMX epoch range based on the $\nu^{-2}$ dependence of the ToAs that fall within that epoch range.
These DMX model parameters are fitted simultaneously together with the rest of the timing model free parameters.
Note that we do not fit for the overall DM simultaneously with the DMX parameters as they are covariant with each other.

To compare and contrast the results from WT, we also used {\tt DMcalc} \citep{kmj21} to obtain the DMs at each epoch. \texttt{DMcalc} is a script written to automate many of the steps in obtaining DM from each epoch using the \texttt{PSRCHIVE} Python interface and \texttt{TEMPO2}. In this method, we use a high S/N, frequency resolved template to obtain the ToAs and estimate DM using them for every epoch. Huber regression is used to remove the large outlier ToAs before estimating the DM using \texttt{TEMPO2}. We made a high S/N template for each pulsar by using the \texttt{psrsmooth} program of \texttt{PSRCHIVE}. Similar to WT, the data from the same epoch and same channel resolution (64 channels) are used. The data of each of these pulsars are passed through {\tt DMcalc} along with the above created high S/N templates and the parameter files as used in WT method (without the DMX values and after updating the DM value to the one with which the template is aligned). The DM timeseries of each of the epochs is obtained.

The DM estimates from WT, NT, and {\tt DMcalc} have offsets among them. 
J1939$+$2134 has the smallest difference in median DMs, which is $2.7 \times $10$^{-5}$~cm$^{-3}$pc, between WT and {\tt DMcalc}. The maximum offset is seen for J1643--1224, which is $1.4 \times $10$^{-2}$~cm$^{-3}$pc, between WT and {\tt DMcalc}.

We now check how the DM estimates from WT compare with the DM estimates derived from the recently published {\tt DMcalc} method. To establish a correlation (if any) between the general trends in {\tt DMcalc} DM estimates and WT DM estimates, we performed a Spearman's rank correlation test~\citep{astroml}.  The correlation coefficients and $p-$values for each pulsar are listed in Table~\ref{tab:obs}. The $p$-values are computed assuming that the null hypothesis corresponds to no correlation between the pair of datasets.  Since the $p-$values are $< 10^{-2}$, it implies that the DM values between the two measurements are correlated. 
In Figure~\ref{fig:DM}, the median subtracted DM timeseries for WT and {\tt DMcalc} for the five pulsars are shown. It can be seen that the DM precision obtained, in general, is about $\mathcal{O}$(10$^{-4}$)~cm$^{-3}$pc or better. The timing residuals after fitting the selected parameters for each of the pulsars are shown in Figure~\ref{fig:res}. 
The results of the ephemeris fit and comparison between WT and NT methods are consolidated in Table~\ref{tab:par}. In order to obtain a reduced $\chi^2 $ closer to unity, EFAC and DMEFAC parameters were used to model the noise \citep[][]{Alam21}.

The original ToA uncertainties (obtained from the WT analysis) are scaled by the bandwidth-time product $\big(\frac{\delta \nu}{100 MHz} \frac{\tau}{1800 s}\big)^{0.5}$, similar to NG \citep[][]{Alam21}, in order to make a reasonable comparison with the ToA uncertainty reported by NG.

A comparison of the results for each of the pulsars is summarized below.

\subsection{PSR J1643--1224}
For this pulsar, the median DM estimate from the WT analysis is 62.40859~cm$^{-3}$~pc. 
The {\tt DMcalc} median DM estimate is 62.39397~cm$^{-3}$~pc. The DM measurements obtained with these two methods are correlated with a correlation coefficient of 0.67 and $p$-value of 1$\times 10^{-2}$. The median S/N of this pulsar is 343. 

The median scaled ToA uncertainty from WT is 3.58 $\mu$s with a postfit rms of about 0.49 $\mu$s. In comparison, NG reports a median ToA uncertainty of 0.46 $\mu$s at 1.4~GHz. Our precision is a factor of 7.8 lower than NG. 

\subsection{PSR J1713+0747}
For this pulsar, the median DM estimate from WT analysis is 15.98957~cm$^{-3}$~pc. 
The median DM estimate from {\tt DMcalc} is 15.99003~cm$^{-3}$~pc. The DM measurements obtained with these two methods are correlated with a correlation coefficient of 0.58 and $p$-value of 2$\times 10^{-2}$. The median S/N of this pulsar is 178, which is the least in our sample.

The median scaled ToA uncertainty from WT is 0.81 $\mu$s with a postfit rms of about 0.06 $\mu$s.  In contrast, NG reports a median ToA uncertainty of 0.043 $\mu$s at 1.4~GHz. Our precision is a factor of 18.7 lower than NG.  

\subsection{PSR J1909--3744}
For this pulsar, the median DM estimate from WT analysis is 10.39113~cm$^{-3}$~pc. 
The median DM estimate from {\tt DMcalc} is 10.39085~cm$^{-3}$~pc. The DM measurements obtained with these two methods are highly correlated with a correlation coefficient of 0.84 and $p$-value of 8$\times 10^{-5}$. The median S/N for this pulsar is 261. This pulsar has a sharp pulse profile with no scatter broadening. 

The median scaled ToA uncertainty from WT is 0.46 $\mu$s with a postfit rms of about 0.03 $\mu$s.  In contrast, NG reports a median ToA uncertainty of 0.086 $\mu$s at 1.4~GHz. Our precision is a factor of 5.3 lower than NG.

\subsection{PSR J1939+2134}
For this pulsar, the median DM estimate from WT analysis is 71.017317~cm$^{-3}$~pc. 
The median DM estimate from {\tt DMcalc} of 71.017344~cm$^{-3}$~pc. The DM measurements obtained with these two methods are highly correlated with a correlation coefficient of 0.92 and $p$-value of 4$\times 10^{-5}$.  The median S/N for this pulsar is 1175, which is the best in our sample. 

The median scaled ToA uncertainty from WT is 0.39 $\mu$s with a postfit rms of about 0.04 $\mu$s. In contrast, NG reports a median ToA uncertainty of 0.01 $\mu$s at 1.4~GHz. Our precision is a factor of 38.5 lower than NG.

\subsection{PSR J2145--0750}
For this pulsar, the median DM estimate from WT analysis is 8.99820~cm$^{-3}$~pc.
The median DM estimate from {\tt DMcalc} is 9.00315~cm$^{-3}$~pc. The DM measurements obtained with these two methods are highly correlated with a correlation coefficient of 0.82 and $p$-value of 2$\times 10^{-4}$. The median S/N for this pulsar is 851. 

The median scaled ToA uncertainty from WT is 1.22 $\mu$s with a postfit rms of about 0.11 $\mu$s. In comparison, NG reports a median ToA uncertainty of 0.48 $\mu$s at 1.4~GHz. Our precision is a factor of 2.5 lower than NG.

\begin{table*}
\begin{adjustbox}{width=17cm}
\begin{tabular}{|c|c|c|c|c|c|c|c|}
\hline
\hline
\textbf{PSR} & \textbf{ToA} & \textbf{Postfit} & \textbf{Median} &\textbf{Median DM}& \textbf{Lowest DM} & \textbf{{\tt DMcalc}} &\textbf{$p$-value}\\
\textbf{} & \textbf{uncertainty} & \textbf{rms} & \textbf{DM}&\textbf{uncertainty}& \textbf{uncertainty} & \textbf{{\tt PulsePortraiture}} &\\
\textbf{} & \textbf{($\mu$s)} & \textbf{ ($\mu$s)} & \textbf{(cm$^{-3}$~pc)}& \textbf{($\times 10^{-4}$cm$^{-3}$~pc)}& \textbf{($\times 10^{-5}$cm$^{-3}$~pc)} & \textbf{Spearman} &\\
\textbf{} & \textbf{} & \textbf{} & \textbf{}& \textbf{}& \textbf{} & \textbf{Coefficient ($\rho $)} &\\

\hline
\texttt{J1643$-$1224} &1.87 &0.49 &62.40859 & 1.1 &9.4 &0.67& $1\times 10^{-2} $\\
\texttt{J1713$+$0747} &0.42 &0.06 &15.98957 & 0.9 &4.9 &0.58 & $2\times 10^{-2} $\\
\texttt{J1909$-$3744} &0.24 &0.03 &10.39113 & 0.2 &1.2 &0.84 & $8\times 10^{-5} $\\
\texttt{J1939$+$2134} &0.20 &0.04 &71.017317 &  0.03 &0.2 &0.92 & $4\times 10^{-5} $\\
\texttt{J2145$-$0750} &0.64 &0.11 &8.99820 & 0.3 &1.8 &0.82 & $2\times 10^{-4} $\\
\hline
\end{tabular}
\end{adjustbox}
\caption{Summary of the results of the {\tt PulsePortraiture} analysis. Column 2 shows the median of the ToA uncertainties for each pulsar. J1939$+$2134 has the best ToA uncertainty in our sample. Column 3 shows the weighted root-mean-square of post-fit timing residuals. J1909--3744 has the best postfit rms in our sample. Column 4 shows the median of the DM obtained for each pulsar. Column 5 contains the median of all the DM uncertainties. 
Column 6 contains the lowest DM precision for each pulsar (precision refers to the uncertainty in the measurements).  J1939$+$2134 has the best DM precision in the $\mathcal{O}$( $10^{-6}$). Column 7 shows the Spearman's rank correlation coefficient for the comparison between {\tt DMcalc} and {\tt PulsePortraiture}. The last column shows the $p$-value corresponding to the null hypothesis of no correlation.  J1939$+$2134 shows the highest correlation and J1713$+$0747 shows the least correlation between the two methods.}
\label{tab:obs}
\end{table*}

\begin{table*}
\begin{tabular}{|l|c|c|c|c|c|c|c|}
\hline
\hline
\textbf{Parameters}& \textbf{WT}& \textbf{WTU}& \textbf{NT}& \textbf{NTU}&\textbf{NT--WT}& \textbf{(NT--WT)/NTU}& \textbf{WTU/NTU}\\
\hline
\textbf{J1643--1224} & & & & & &dimensionless &dimensionless\\
\hline
\texttt{$F_{0}$ (Hz)} & 216.3733404525 & $4.02\times 10^{-10}$& 216.3733404535 & $1.32\times 10^{-9}$&$9.71\times 10^{-10}$ &0.737 &0.305\\
\texttt{$A_1$ (ls)} & 25.072572 & $6.32\times 10^{-6}$ & 25.072556 & $2.07\times 10^{-5}$ &$-1.60\times 10^{-5}$&0.775 &0.306\\
\texttt{$P_b$ (d)} &147.01736 & $3.22\times 10^{-5}$ & 147.01726 & $1.03\times 10^{-4}$ &$-9.51\times 10^{-5}$&0.922 &0.313\\
\hline
Reduced ${\chi}^2$ &1.03 & & & &\\
Dof &6 & & & &\\
\hline
\hline
\textbf{J1713$+$0747} & & & &\\
\hline
\texttt{$F_{0}$ (Hz)} &218.811843784 & $7.30\times 10^{-9}$ &218.811843781 & $8.02\times 10^{-9}$ &$-3.64\times 10^{-9}$ &0.454 &0.911\\
\texttt{$F_{1}$ (Hz/s)} &$-4.01\times 10^{-16}$& $2.36\times 10^{-17}$ &$-3.89\times 10^{-16}$& $2.60\times 10^{-17}$ &$1.19\times 10^{-17}$&0.458 &0.910 \\
\texttt{$A_1$ (ls)} &32.3424251 & $1.04\times 10^{-6}$&32.3424270 & $1.11\times 10^{-6}$&$1.91\times 10^{-6}$&1.730 &0.945\\
\texttt{$P_b$ (d)} &67.8251289 & $6.03\times 10^{-7}$ &67.8251293 & $5.72\times 10^{-7}$ &$3.73\times 10^{-7}$ &0.652 &1.054\\
\hline
Reduced ${\chi}^2$ &1.01 & & & &\\
Dof &9 & & & &\\
\hline
\hline
\textbf{J1909$-$3744} & & & &\\
\hline
\texttt{$F_{0}$ (Hz)} &339.315692407 & $6.08\times 10^{-9}$ &339.315692396 & $4.52\times 10^{-9}$ &$-1.11\times 10^{-8}$ &2.455 &1.346\\
\texttt{$F_{1}$ (Hz/s)} &$-1.66\times 10^{-15}$& $1.95\times 10^{-17}$ &$-1.62\times 10^{-15}$& $1.45\times 10^{-17}$ &$3.53\times 10^{-17}$ &2.440 &1.346\\
\texttt{$A_1$ (ls)} & 1.8979914 & $6.85\times 10^{-7}$&1.8979907 & $5.20\times 10^{-7}$&$-7.04\times 10^{-7}$ &1.354 &1.316\\
\texttt{$P_b$ (d)} &1.533449455 & $2.80\times 10^{-9}$&1.533449442 & $1.92\times 10^{-9}$&$-1.32\times 10^{-8}$&6.876 &1.459\\
\hline
Reduced ${\chi}^2$ &1.07 & & & &\\
Dof &9 & & & &\\
\hline
\hline
\textbf{J1939$+$2134} & & & &\\
\hline
\texttt{$F_{0}$ (Hz)} &641.9282322429 & $9.82\times 10^{-9}$&641.9282322461 & $9.85\times 10^{-9}$&$3.15\times 10^{-9}$ &0.320 &0.997\\
\texttt{$F_{1}$ (Hz/s)} &$-4.31\times 10^{-14}$& $3.12\times 10^{-17}$ &$-4.32\times 10^{-14}$& $3.13\times 10^{-17}$ &$-1.08\times 10^{-17}$ &0.346 &0.997\\
\hline
Reduced ${\chi}^2$ &1.02 & & & &\\
Dof &9 & & & &\\
\hline
\hline
\textbf{J2145$-$0750} & & & &\\
\hline
\texttt{$F_{0}$ (Hz)} &62.2958888011 &$2.50\times 10^{-9}$&62.2958887957 & $3.14\times 10^{-9}$&$-5.42\times 10^{-9}$ &1.726 &0.797\\
\texttt{$F_{1}$ (Hz/s)} &$-1.28\times 10^{-16}$& $7.97\times 10^{-18}$ &$-1.10\times 10^{-16}$& $9.99\times 10^{-18}$ &$1.72\times 10^{-17}$ &1.725 &0.798\\
\texttt{$A_1$ (ls)} &10.1641097 & $1.43\times 10^{-6}$&10.1641082 & $1.72\times 10^{-6}$&$-1.46\times 10^{-6}$ &0.851 &0.833\\
\texttt{$P_b$ (d)} &6.838902519 & $1.85\times 10^{-8}$&6.838902502 & $2.26\times 10^{-8}$&$-1.77\times 10^{-8}$&0.782 &0.816\\
\hline
Reduced ${\chi}^2$ &1.00 & & & &\\
Dof &9 & & & &\\
\hline
\hline

\end{tabular}
\caption{Comparison of the WT and NT postfit parameters of the pulsars in our sample are tabulated here. WTU is the wideband timing uncertainty and NTU is the narrowband timing uncertainty. $F_{0}$ is the pulsar rotation frequency, $F_{1}$ is the pulsar rotation frequency first derivative, $A_1$ is the projected pulsar semi-major axis in light seconds (ls), $P_b$ is the period of the binary orbit, and degrees of freedom (dof) is the sum of the number of ToAs and the number of DM measurements from the wideband timing from which the timing parameters are subtracted which include DMX parameters. In the seventh column, the absolute difference is divided by NTU.}
\label{tab:par}
\end{table*}

\section{Conclusions and Discussions}
\label{sec:conc}
In this work, we have demonstrated the application of wideband timing using {\tt PulsePortraiture} on low-frequency (300--500 MHz) data for  five millisecond pulsars: PSRs J1643--1224, J1713$+$0747, J1909--3744, J1939$+$2134, and J2145--0750, observed at uGMRT as part of the InPTA program. These pulsars show different morphologies in pulse shapes and varying degrees of broadening in their pulse profiles.  DM estimates 
with this method are consistent with techniques, such as {\tt DMcalc} \citep{kmj21}, which  use data with narrow sub-bands. At the same time, this technique  simultaneously provides high precision ToAs. PCA analysis, employed for this technique, indicates that we require a minimum of three eigenprofiles for PSRs J1643$-$1224, J1713+0747, and J1939+2134; and  two eigenprofiles for PSRs J1909$-$3744 and J2145$-$0750 to capture the profile evolution with frequency. We obtained DM precision ranging between 3 $\times$ 10$^{-6}$ ~cm$^{-3}$pc for PSR J1939+2134 to 1 $\times$ 10$^{-4}$ ~cm$^{-3}$pc for PSR J1643--1224. Using this method, we get sub-microsecond post-fit average residuals. We achieved the best post-fit residuals of about 30 ns for PSR J1909$-$3744.  

Using the dispersion formula 
\begin{equation*}
\Delta t = 4.148808 \text{ ms} \times \bigg[ \bigg(\frac{f_{lo}}{\text{GHz}}\bigg)^{-2} - \bigg(\frac{f_{hi}}{\text{GHz}}\bigg)^{-2}\bigg] \times \bigg(\frac{DM}{\text{cm}^{-3}\text{pc}}\bigg),
\end{equation*}
\citep[See Appendix A 2.4][]{2012hpa..book.....L}, it can be shown that the precision in DM measurements obtained over our 200 MHz bandwidth (e.g., 300$-$500 MHz) of  $\mathcal{O}$(10$^{-5}$) is at least an order of magnitude better than that over a wide high frequency band (e.g., 700$-$4000 MHz), which is  $\mathcal{O}$(10$^{-4}$) ~cm$^{-3}$pc (for assumed typical 1 $\mu$s ToA errors). These 300$-$500 MHz observations provide a S/N comparable to the GHz bandwidth observations at high frequencies, as pulsars are much brighter at 400 MHz. In addition, our results show that the application of WT to our band can provide post-fit residuals comparable to high frequency data by taking care of ISM effects considerably. Thus, WT of such low frequency observations is  capable of providing not only more accurate DM estimates, but also high precision ToAs directly. It will be interesting to make a direct comparison between the analysis of low and high frequency PTA data in a future IPTA data combination to investigate this further. 

We compare these low-frequency ToA residuals and DM uncertainties with the results published in the literature for the same pulsars, both at low \citep[{\tt DMcalc: }][]{kmj21} and high frequencies  \citep[NG: ][]{Alam21}.
In the low frequency band, our DM estimates show a strong correlation with the results from {\tt DMcalc}. %
Now, WT technique has considerable advantages over the traditional timing techniques. Firstly, WT is more amenable to automation with a one-step analysis. In contrast, analysis methods using sub-bands, such as the traditional narrow-band analysis or {\tt DMcalc}, require a multi-step iterative approach with the DM estimation followed by timing in an iterative loop. Secondly, traditional analysis either ignores profile evolution or pulse broadening or at best approximates it. In contrast, WT incorporates this as an essential ingredient of analysis. In Figure \ref{fig:uncert}, we compare the uncertainties from both the methods for these five pulsars. With the exception of PSR J1643-1224 and PSR J1939+2134, the uncertainty from both the methods are consistent. The inconsistency could be related to the fact that we have only considered profile evolution and not scattering effects, and these pulsars show significant pulse broadening at low frequencies. We plan to investigate this further in a future work. Lastly, the WT technique utilizes the S/N of the  entire wideband observations to provide high precision ToA unlike the narrow-band technique. This also results in a single high S/N band- averaged ToA rather than 16 to 32 lower S/N ToAs. This significantly reduces the dimensionality of subsequent Bayesian analysis, which is employed for the detection of GWs. Thus, the consistency of the DM estimates between the  WT and traditional methods provides support for a preferential use of WT technique at low frequencies, in particular, and hint at an increasing reliance on WT technique for future PTA and IPTA data release, in general.

A comparison of the timing solutions obtained from WT and traditional NT are presented in Table \ref{tab:par}. As is evident from the aforementioned table, WT produces timing solutions consistent with NT, with typical uncertainties in fitted parameters smaller than NT.

In our application, the pulse broadening was assumed to be stable over the observation epochs. This may not be the case for all the pulsars. An example is PSR J1643$-$1224, where variable pulse broadening at a given frequency was reported earlier \citep{slk+16}. Epoch to epoch variation of the profile evolution with frequency has also been reported in PSR J1713+0747 \citep{ssj+21}. An extension of WT to include such a variation will be interesting and is planned in future. Similar extension to combine widely separated multiple bands is also planned in future. 

A comparison with the median ToA uncertainties at  high frequencies, such as those obtained by NG at 1.4 GHz, indicates that our ToA uncertainties are of the same order (2.5 to 7.8 times), except for two pulsars (PSRs J1713$+$0747 J1939$+$2134). These findings suggest that low frequency data, analysed with WT technique, can provide a precision similar to high frequency data for gravitational wave detection experiments. Given the steep spectrum of radio pulsars, this not only enables high precision measurements with smaller observation duration per pulsar at low frequency (as pulsars are much brighter at these frequencies), but also a higher cadence than currently employed with the same telescope time. Additionally, several weaker MSPs can be included in the PTA ensemble. Not only this can provide a more uniform sky coverage for useful sampling of the Hellings and Downs overlap reduction function \citep{hellings1983upper}, but also significantly increase the sensitivity \citep{Siemens_2013} to the stochastic gravitational wave background. Thus, our results suggest that wideband low frequency observations can play at least an equal, if not better role, in PTA experiments.

With the Square Kilometer Array \citep[SKA:][]{Carilli:2004nx} telescope becoming available in the near future, wideband observations with SKA-low (200$-$350 MHz) and SKA-mid (350$-$1000 MHz) promise to provide high quality data not only for nanoHertz gravitational wave discovery, but also for post discovery gravitational wave science. Wideband techniques are likely to play a very important role in analysis of these data from SKA.

{\bf Note Added:} After this paper appeared on arXiv, another work also  applied the {\tt PulsePortraiture} based wideband timing analysis to eight millisecond pulsars observed with the uGMRT~\citep{Sharma2022}. This work also finds better DM and timing precision compared to the narrow-band method in accord with our results. \newline

Software: {\tt matplotlib} \citep{4160265}, {\tt PSRCHIVE} \citep{2010PASA...27..104V}, {\tt PulsePortraiture} \citep{2014ApJ...790...93P}, {\tt TEMPO} \citep{2015ascl.soft09002N}, {\tt TEMPO2} \citep{2006MNRAS.369..655H}, {\tt DMcalc} \citep{kmj21}, {\tt RFIClean} \citep{Maan+2020}, {\tt pinta} \citep{pinta}\newline

Facility: uGMRT.\newline

\section*{Acknowledgements}
We are grateful to the anonymous referee for very useful and constructive feedback on our manuscript.
This work is carried out by InPTA, which is part of the International Pulsar Timing Array consortium. We thank the staff of the GMRT who made our observations possible. GMRT is run by the National Centre for Radio Astrophysics of the Tata Institute of Fundamental Research.  BCJ, PR, AS, SD, LD, and YG acknowledge the support of the Department of Atomic Energy, Government  of India, under project identification \# RTI4002.  BCJ and YG acknowledge support from the Department of Atomic Energy, Government of India, under project \# 12-R\&D-TFR-5.02-0700.  MPS acknowledges funding from the European Research Council (ERC) under the European Union’s Horizon 2020 research and innovation programme (grant agreement No. 694745).  AC acknowledge support from the Women’s Scientist scheme (WOS-A), Department of Science \& Technology, India.  NDB acknowledge support from the Department of Science \& Technology, Government of India, grant SR/WOS-A/PM-1031/2014. SH is supported by JSPS KAKENHI Grant Number 20J20509. KT is partially supported by JSPS KAKENHI Grant Numbers 20H00180, 21H01130 and 21H04467, Bilateral Joint Research Projects of JSPS, and the ISM Cooperative Research Program (2021-ISMCRP-2017).

\section*{Data availability}
The data underlying this article will be shared on reasonable request to the corresponding author.




\bibliographystyle{mnras}
\bibliography{wideband} 




\appendix


\bsp	
\label{lastpage}
\end{document}